\documentclass[reprint, amsmath,amssymb, aps,prl,superscriptaddress]{revtex4-2}

\usepackage{graphicx}
\usepackage{dcolumn}
\usepackage{bm}
\usepackage[margin=2.0cm]{geometry}
\usepackage{amsmath}
\usepackage{braket}
\usepackage{appendix}
\usepackage{float}
\usepackage{siunitx}
\usepackage{amssymb}
\usepackage{color}
\usepackage{multirow}
\usepackage{longtable}
\usepackage{ulem}

\begin{document}



\title{Universality in odd-even harmonic generation \\ and application in terahertz waveform sampling}

\author{Doan-An Trieu}%
\affiliation{Computational Physics Key Laboratory K002, Department of Physics,\\Ho Chi Minh City University of Education, Ho Chi Minh City 72711, Vietnam}%

 \author{Ngoc-Loan Phan}%
 \email{loanptn@hcmue.edu.vn}
 \affiliation{Computational Physics Key Laboratory K002, Department of Physics,\\Ho Chi Minh City University of Education, Ho Chi Minh City 72711, Vietnam}%

 \author{Quan-Hao Truong}%
\affiliation{Computational Physics Key Laboratory K002, Department of Physics,\\Ho Chi Minh City University of Education, Ho Chi Minh City 72711, Vietnam}%


\author{Hien T. Nguyen}%
\affiliation{Tay Nguyen University, Buon Ma Thuot City 63161, Vietnam}%

\author{Cam-Tu Le}%
\affiliation{Atomic Molecular and Optical Physics Research Group, Advanced Institute of Materials Science, Ton Duc Thang University, Ho Chi Minh City 72915, Vietnam}%

\affiliation{Faculty of Applied Sciences, Ton Duc Thang University, Ho Chi Minh City 72915, Vietnam}%

\author{DinhDuy Vu}%
\affiliation{Computational Physics Key Laboratory K002, Department of Physics,\\Ho Chi Minh City University of Education, Ho Chi Minh City 72711, Vietnam}%


\author{Van-Hoang Le}%
\email{hoanglv@hcmue.edu.vn}
\affiliation{Computational Physics Key Laboratory K002, Department of Physics,\\Ho Chi Minh City University of Education, Ho Chi Minh City 72711, Vietnam}%

\date{\today}

\begin{abstract}
Odd-even harmonic generation contains the system's dynamical symmetry breaking, and its decoding is desirable not only in understanding physics but also for applications of dynamics extraction. In this work, we discover a simple universal relation between the odd-even harmonics and the asymmetry of the THz-assisted laser-atomic system -- atoms in a primary midinfrared laser pulse combined with a THz laser. We demonstrate numerically and then analytically derive the dependence of the harmonic even-to-odd ratio on the THz electric field. Notably, the functional form of this dependency reveals a universal scaling independent of the parameters of both the primary pulse and atomic target. This universality inspires us to propose a pump-probe scheme for THz waveform sampling from the even-to-odd ratio, accessible from conventional compact setups.
\end{abstract}

\maketitle

\textit{Introduction ---} Recently, high-order harmonic generation (HHG) from an asymmetric laser-target system has gained great attention since it reveals deeper structures in HHG spectra~\cite{Frumker1:prl12,Chen:prl13,Pedatzur:natphys15,Kraus:NatCom15,Chen:epl19,Li:pra19,Yin:apl21,Nguyen:pra22,He:NatCom22,Ngan:pra22,Wang:pra17}. The most prominent feature in such symmetry-breaking systems is the emergence of even harmonics, which apparently separate them from symmetric counterparts, which emit only odd harmonics~\cite{Ben-Tal:jpb93}. Therefore, the odd-even pattern in HHG spectra must encode the dynamical asymmetry information of the laser-target system~\cite{Frumker1:prl12,Chen:prl13,Pedatzur:natphys15,Yin:apl21,Nguyen:pra22,Kraus:NatCom15,Chen:epl19,Li:pra19,He:NatCom22,Ngan:pra22}, and thus a universal relation between them can be helpful in many extraction applications.

One route to break the spatial-temporal symmetry is adding a weaker external static electric field to the primary multi-cycle laser pulse \cite{Bao:pra96,Wang:jpb98,Hong:jpb07,Zhao:njp11,Le:PCCP22,Silaev:ol22}. However, to yield a visible effect, the static field needs to be impractically strong ($\sim$MV/cm). Nevertheless, with recent developments of powerful terahertz (THz) sources, a quasi-static strong electric field can be engineered into the HHG process \cite{Hong:oe09,Balogh:pra11,Jia:oe15,Ge:oe15,Babushkin:pra22,Milosevic:ol22,Brennecke:prl22,Silaev:ol22,Silaev:jpcs22,Mao:prb22}. This THz-assisted HHG has been studied mostly in the harmonic conversion efficiency or the plateau structure~\cite{Hong:oe09,Balogh:pra11,Jia:oe15,Ge:oe15,Babushkin:pra22} but less in other notable aspects such as odd-even harmonic spectra. Some initial estimates have been made~\cite{Wang:jpb98,Silaev:jpcs22}, but a direct relation between the odd-even pattern and the asymmetricity of the THz-assisted laser-atomic system has not yet been established. This relation is not only of intellectual interest but can be critical to many applications such as extracting quantum dynamics inside atoms/molecules, manipulating electron trajectories on the attosecond time scale, or in THz waveform sampling, which has been gaining attention recently~\cite{Pedatzur:natphys15,Ge:oe15,Zhang:NatPho17,Zhang:FronOpt21,Silaev:ol22}.

\begin{figure*}[!htb]
\begin{center}
\includegraphics[width=0.85\linewidth]{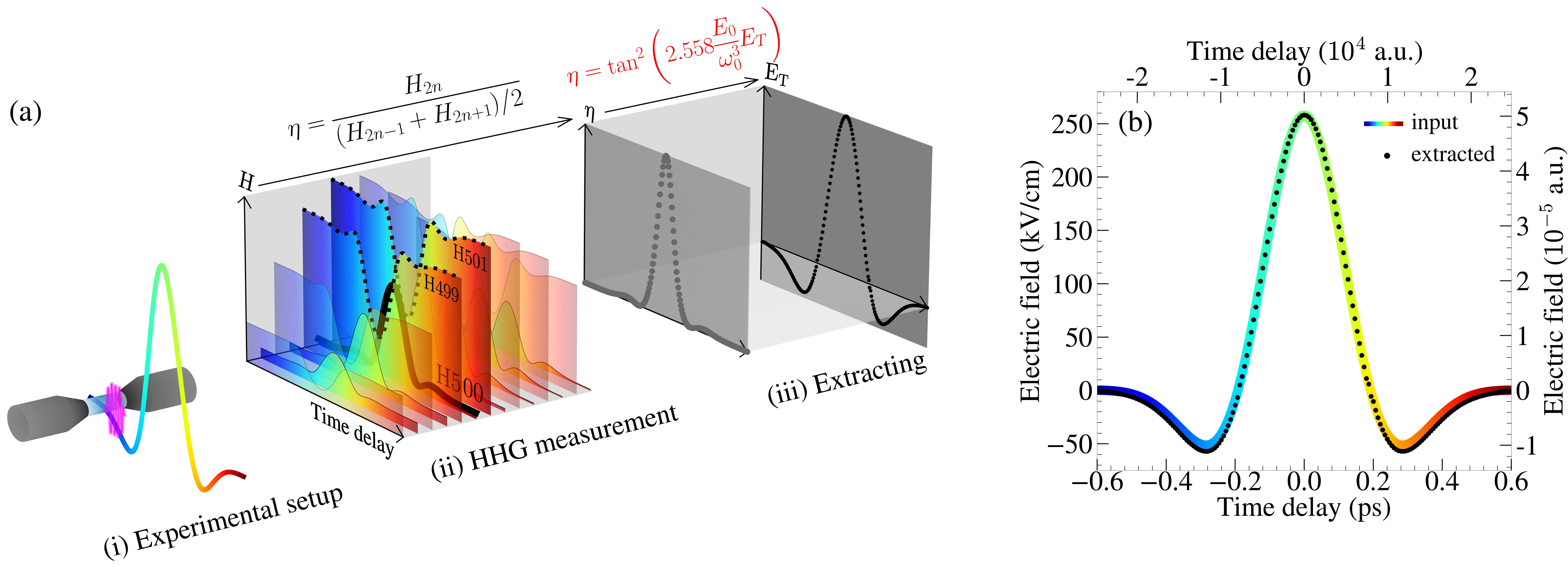}
\end{center}
\caption{(a) Pump-probe scheme for sampling THz waveforms with sequential steps: (i)~experimental setup with the THz pulse (pump) and a primary mid-IR laser (probe) both shinning on a gas jet; (ii)~measurements of intensities of even and two adjacent odd harmonics at the cutoff region with pump-probe time delays; (iii)~extracting step with the first computing the even-to-odd ratio from HHG signals and then extracting THz electric field using the universal rule. (b)~Comparison between the extracted (dotted curve) and benchmark (solid curve) THz pulses. Five-cycle trapezoidal primary pulses with the intensity of $2.5\times 10^{14}$~W/cm$^2$ and wavelength of 2000~nm are used as probe lasers. }
\label{fig:extract}
\end{figure*}

In this work, we accomplish two goals. First, we thoroughly study the response of the odd-even HHG to the variation of the THz electric field in a THz-assisted laser-atomic system. The purpose is to achieve a quantitative connection between the measurable harmonic even-to-odd ratio and the symmetry-breaking THz electric field. Interestingly, we discover a universal and simple rule for the even-to-odd ratio as a function of the scaled THz electric field. This finding stimulates us to proceed with the second goal of proposing a general method for sampling the THz pulse waveform. Our method can retrieve the time-resolved waveform with high accuracy from the harmonic even-to-odd ratio by an algorithm shown in Fig.~\ref{fig:extract}. In short, we collect the HHG signals emitted from the gas jet inside the intersection between a primary midinfrared (mid-IR) linearly polarized laser and the THz pulse. Our algorithm allows a direct reconstruction of the instantaneous THz strength through the even-to-odd ratio of harmonics of the HHG readout. By tuning the pump-probe time delay, we can profile the entire THz pulse. In this vein, we note that waveform detection is another important area of THz science that is intensively investigated in addition to its generation~\cite{Zhang:NatPho17,Zhang:FronOpt21,Silaev:ol22}. Within currently available methods, active matter has to be carefully chosen; thus, finding a new temporal detection scheme free from external parameters is experimentally meaningful.

\begin{figure}[!htb]
	\begin{center}
		\includegraphics[width=0.85\linewidth]{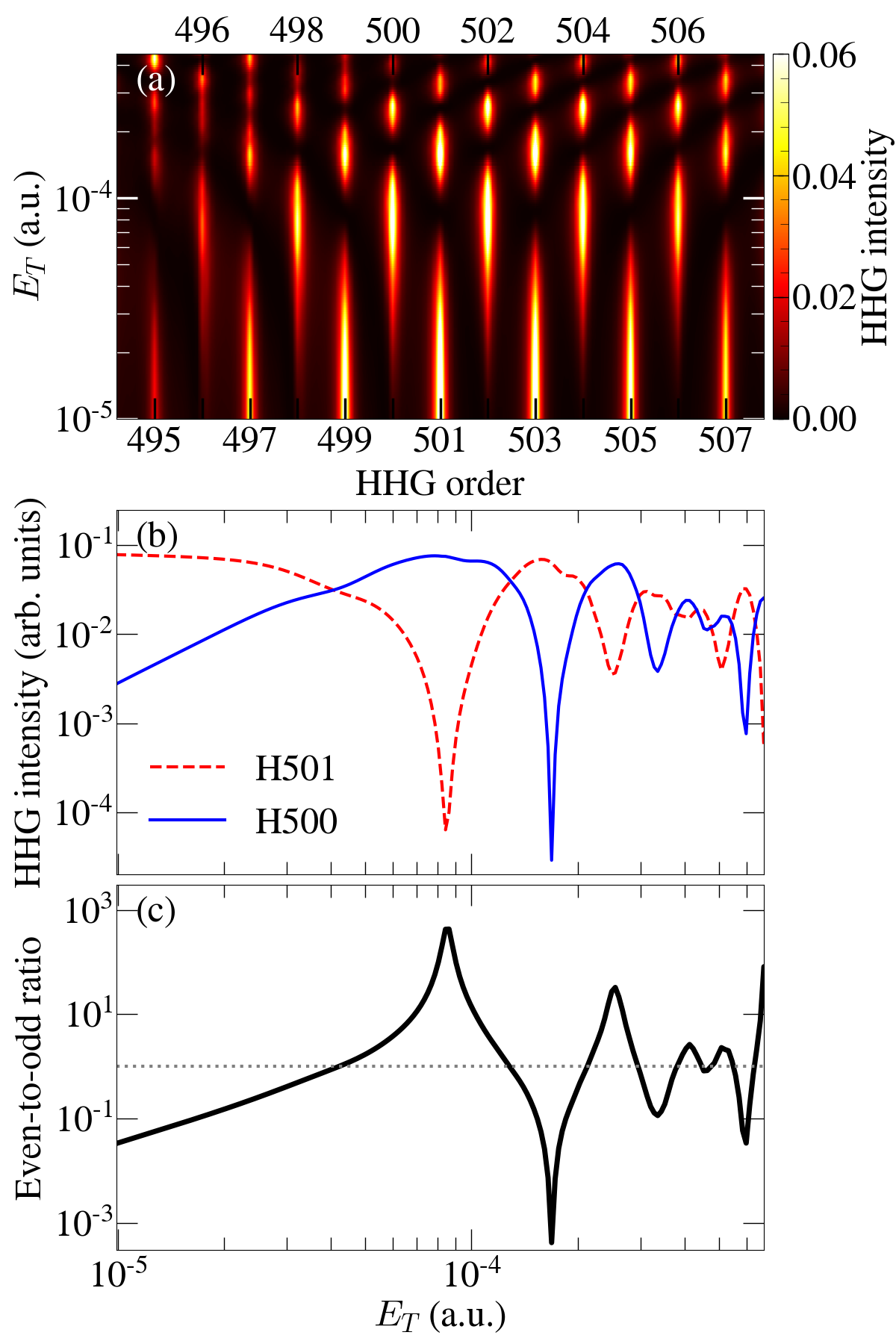}
	\end{center}
	\caption{Response to changing of THz electric field for (a)~Intensity of resolved odd-even harmonic spectra, (b)~Selected odd (H501) and even (H500) harmonics at the cutoff, and (c)~Harmonic even-to-odd ratio. The primary mid-IR laser pulse with the same parameters as in Fig.~\ref{fig:extract} is used for the HHG process; the THz field with a frequency of 1.3~THz (231~$\mu$m) is used; the color bar in Panel~(a) decodes the HHG intensity in arbitrary units; the dotted horizontal line in Panel~(c) shows the unity.
	}
\label{fig:odd-even}
\end{figure}

{\textit{Universal response of even-to-odd ratio to THz electric field} --- } We first look thoroughly into the response of resolved odd-even harmonics emitted from a hydrogen atom in the combined linearly polarized primary mid-IR laser pulse and the slow-varying THz field, as shown in Fig.~\ref{fig:odd-even}(a). Here, the HHG data are calculated by numerically solving the time-dependent Schr\"{o}dinger equation (TDSE method)~\cite{suppl}. Since electron wave packets are mostly spread along the laser polarization direction, the one-dimensional model is sufficient for the HHG study as it requires low computational overhead but is quantitatively comparable to the three-dimensional model~\cite{Chili:pra10,Majorosi:pra18}.  We consider an example of a mid-IR pulse with a 10-cycle trapezoidal envelope, the intensity of $2.5\times 10^{14}$~W/cm$^2$, and the wavelength of 2000~nm. It is clear that when the added THz field $E_T$ is much weaker than the laser field [$< 10^{-5}$~a.u. (51~kV/cm)], the spectrum is pure-odd harmonics~\cite{Ben-Tal:jpb93}. Upon increasing the THz strength, the space-time symmetry (half-period-time translation combined with spatial inversion) of the original system is violated, leading to the emergence of even harmonics.

For a more focused picture, Fig.~\ref{fig:odd-even}(b) shows the response of harmonic intensity with the THz strength for the two selected even (500${\mathrm{th}}$) and odd (501${\mathrm{st}}$) harmonics at the cutoff, denoted respectively as H500 and H501. Their behavior can be visually partitioned into two regions. (i) For the first region, the intensity of the even harmonics gradually increases, while that of the odd harmonics is almost unchanged. (ii) After the first intersection at $E_T \approx 4 \times 10^{-5}$~a.u. ($\sim 0.2$~MV/cm) that indicates the equal presence of the even and odd harmonics, the second region begins with the intensity fluctuation of the odd and even harmonics around their average values. Notably, the odd and even harmonics at some specific THz field values alternatively undergo deep minimums corresponding to pure-even or pure-odd spectra, as illustrated in Fig.~\ref{fig:odd-even}(a).

\begin{figure} [!htb] 
    \begin{center}
    \includegraphics[width=0.85\linewidth]{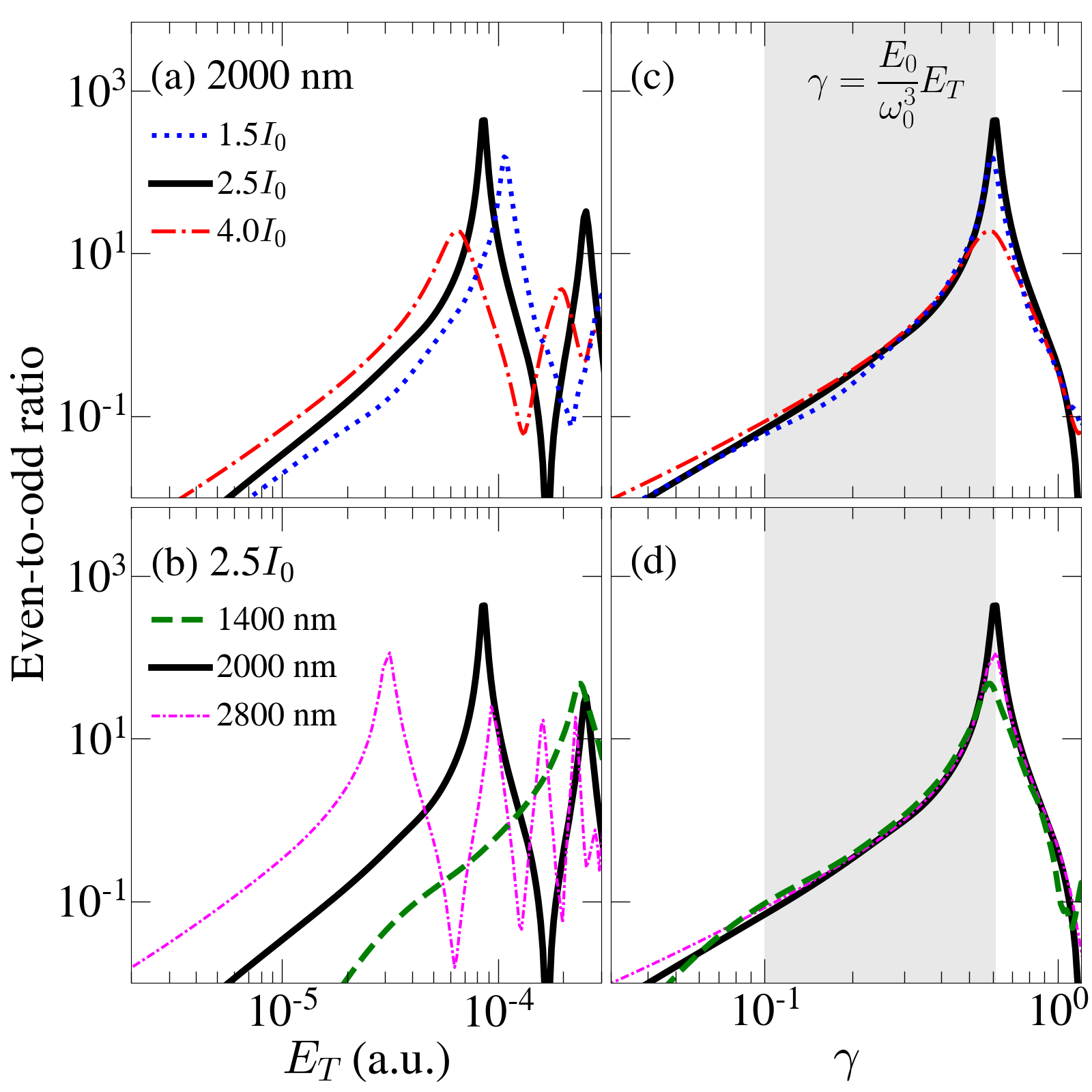}
    \end{center}
    \caption{Dependence of harmonic even-to-odd ratio on pure (left panels) and scaled (right panels) THz electric field for a hydrogen atom in various THz-assisted primary laser pulses with (a) different intensities (fixed wavelength of 2000~nm) and (b) different wavelengths (fixed intensity of 2.5$I_0$), where $I_0 = 1\times$10$^{14}$~W/cm$^2$. The grey-shaded areas in Panels~(c) and (d) highlight the region of stable even-to-odd ratio.
 }
    \label{fig:universal}
\end{figure}

Since the absolute harmonic efficiency is strongly affected by the laser field, we introduce a dimensionless quantity by taking the intensity ratio between the even harmonic and the average of the two adjacent odd ones, which is referred to as the even-to-odd ratio. The variation of this quantity with respect to the THz field is shown in Fig.~\ref{fig:odd-even}(c). However, taking the intensity ratio does not entirely cancel the effect of the primary laser still on the THz-dependent even-to-odd ratio, as can be seen in the sensitivity to different laser intensities and wavelengths shown in Figs.~\ref{fig:universal}(a) and ~\ref{fig:universal}(b). Specifically, the sensitivity to laser wavelength~[Fig.~\ref{fig:universal}(b)] is much more severe than laser intensity~[Fig.~\ref{fig:universal}(a)]. Remarkably, we find that by rescaling the THz electric field $E_T$ by the factor $E_0/\omega_0^3$ as
\begin{equation}
 \gamma = \dfrac{E_0}{\omega_0^3} E_T,
\end{equation}
where $E_0$ and $\omega_0$ are the primary laser's peak amplitude and carrier frequency, the data collapse quantitatively within the range $\gamma \gtrsim 0.1$ (see Fig.~\ref{fig:universal}). Here, the region $\gamma$ below $0.1$ suffers from significant numerical noise. We also numerically verify the data collapse with respect to the duration (5, 10, 15, 20 cycles) of the primary laser pulse and, more interestingly, with different active atomic targets (H, He, Ne, Ar)~\cite{suppl}. Furthermore, the universal response as a function $\gamma$ is observed not only for harmonics at the cutoff, but also for those below the cutoff with an additional condition of good phase matching in HHG experiments~\cite{suppl}.

{\textit{Analytical formula} --- } The universal relation between the harmonic even-to-odd ratio and scaled THz electric field observed numerically above motivates us to uncover its underlying physics. We show that it is not accidental but can be proven rigorously by the quantum-orbit theory~\cite{Salieres:Sci01}. Within this framework, the generation of harmonic order $N$ is the coherent interference of attosecond bursts emitted each half of an optical cycle: $H(N\omega_0) \approx |D_1 e^{-i\Phi_1} + D_2 e^{-i\Phi_2}|^2$, where $D_{1(2)}$ and $\Phi_{1(2)}$  are their amplitude and phase. Therefore, the even-to-odd ratio is mostly controlled by the relative intensity $|D_2|^2/|D_1|^2$ and phase difference $\Delta \Phi$ between attosecond bursts, which in turn is associated with the difference of quasi-classical action $\Delta S$ of free electrons in the laser electric field as $\Delta \Phi = N \pi - \mathrm{Re}(\Delta S)$. Adding a perturbative THz field does not change significantly the intensity of attosecond bursts, i.e., $|D_2|^2/|D_1|^2 \approx 1$. However, it distorts the quasi-classical action by $S_T   =  C \gamma$ to the first order, leading to the change in the phase of attosecond bursts, which in turn modifies the even-to-odd ratio as
\begin{equation}
\eta = \tan^2 \left( C \gamma \right).
\label{eq:e2o}
\end{equation}
 Here, $C = 2 \sin \theta ~(\Delta \theta \cos \Delta \theta  - \sin \Delta \theta  )$ with $\theta = \omega_0 (t_r+t_i)/2$ and $\Delta \theta = \omega_0 (t_r - t_i)/2 $ is a real dimensionless coefficient depending on the instants of harmonic ionization $t_i$ and recombination $t_r$. We emphasize that $C$ is a constant for each harmonic energy, and does not contain any parameters of the laser or target, either explicitly or implicitly. Specifically for harmonics at the cutoff, $C \approx 2.558$.
See more detailed calculations in Supplementary~\cite{suppl}.

The analytical formula (\ref{eq:e2o}) demonstrates a direct connection between the even-to-odd ratio $\eta$ (a normalized quantity characterizing the asymmetry of measurable output) and the dimensionless scaled THz electric field~$\gamma$ (the normalized symmetry-breaking factor). Most importantly, Eq.~(\ref{eq:e2o}) is free from parameters of the primary laser pulse parameters (except implicitly in $\gamma$) and the atomic target, implying the universality of the response of harmonic even-to-odd ratio to the scaled THz field. Besides, this analytical expression also shows a periodic modulation of the harmonic even-to-odd ratio with the period of $\pi/C$. It alternatively undergoes maxima and minima, generating instants of the pure-even and pure-odd harmonic spectra. Based on the relation~(\ref{eq:e2o}), we further refer to the dimensionless quantity $\gamma$ as a parameter describing the asymmetric degree of the laser-target system and call it the asymmetry parameter.

\begin{figure} [htb!]
	\begin{center}		\includegraphics[width=1\linewidth]{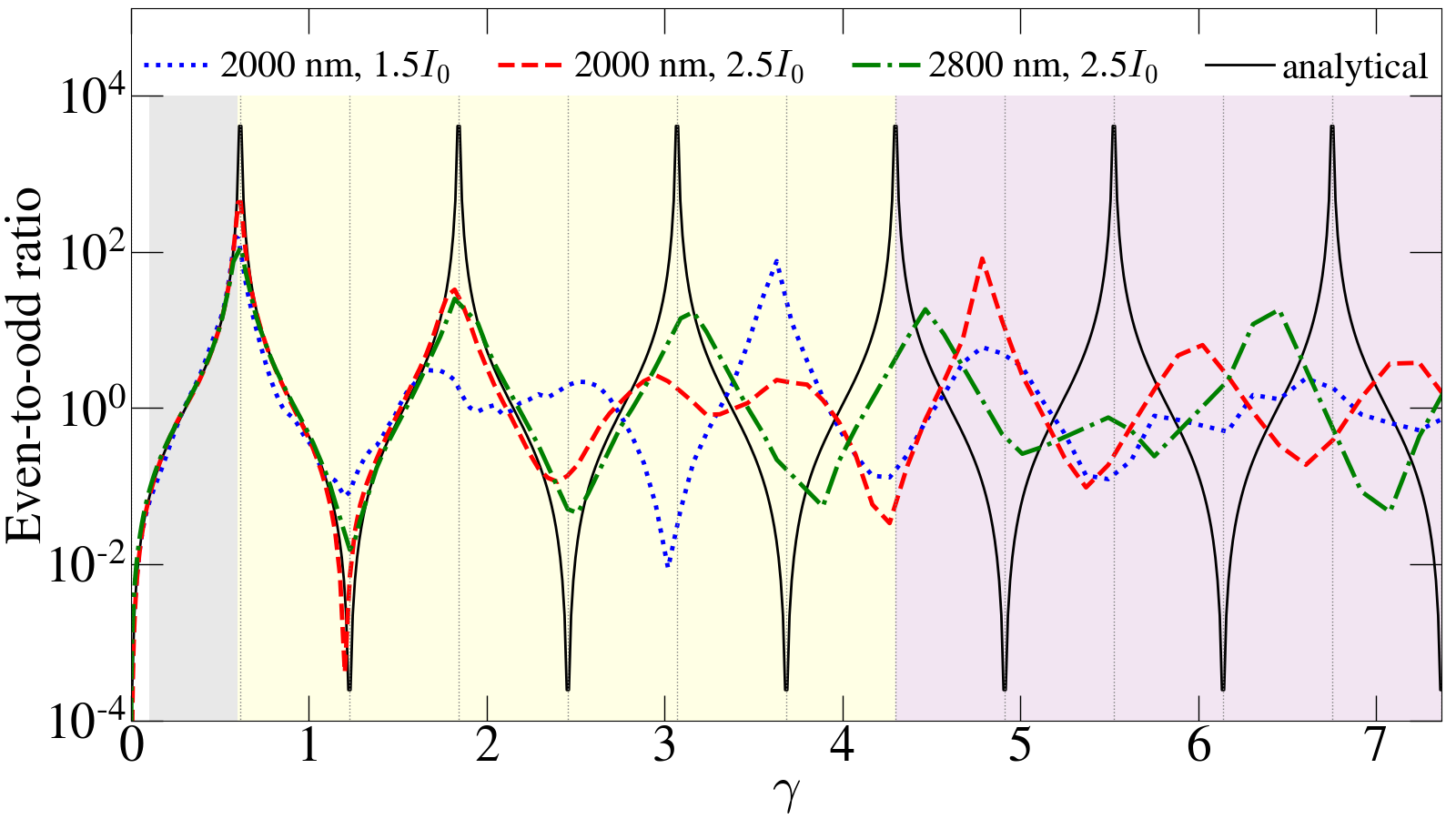}
	\end{center}
	\caption{Response of harmonic even-to-odd ratio to a wide range of scaled THz fields calculated by the analytical formula $\eta=\tan^2(2.558\, \gamma)$ (black solid curve) and numerical TDSE method (dashed and dotted color curves) for harmonics at the cutoff using different primary lasers enclosed in the legend. The grey-, cream-, and mauve-shaded areas cover three regions with different underlying physics mechanisms.
	}
	\label{fig:class}
\end{figure}

Figure~\ref{fig:class} shows an overall visualization of the even-to-odd ratio versus the asymmetry parameter $\gamma$ for the harmonics at the cutoff. This visualization is based on the data obtained in two different ways, using analytical relations and direct numerical calculations by the TDSE method. Comparing the results of the two methods reveals three regimes with different levels of analytic-numeric agreement, implying various physical mechanisms.
(i)~For a small asymmetry parameter \mbox{$ 0.1 \lesssim \gamma \lesssim 0.6 $} (gray-shaded area), the analytical formula quantitatively predicts the even-to-odd ratio obtained from numerically solving the TDSE with various primary laser parameters and atomic targets. The agreement is expected within perturbative strengths of the THz field because it only modifies the electron quasi-classical motion in the continuum energy region but does not affect the ionization and recombination steps. 
(ii)~For higher asymmetric parameters \mbox{$ 0.6 \lesssim \gamma \lesssim 3.5\pi/C \equiv 4.3$} (cream-shaded area), the relation (\ref{eq:e2o}) fails to match the magnitude. Moreover, the disparity caused by the laser's parameters becomes prominent, unlike the perturbative regime. However, the reversal points (from $>1$ to $<1$, and vice versa) still fall into the sequence of $(k+0.5)\pi/2C$ with $k\in \mathbb{N}$. The main reason is moderately strong THz field participates in the ionization step beyond distorting the electron quasi-classical motion. This induces the imbalance between adjacent attosecond bursts, thus reducing the interference contrast and causing irregular modulation of the even-to-odd ratio.
(iii)~With $\gamma \gtrsim 4.3$ (mauve-shaded area), the numerical even-to-odd ratio becomes highly disordered, and the classical description~(\ref{eq:e2o}) fails to predict both the magnitude and spacing between the reversal points. Here, the intense THz field modifies ionized electrons' travel time compared to the field-free case. Additionally, intense THz fields dominate over the primary laser field in driving electron trajectories, thus altering the plateau structure of HHG spectra. 

In short, universality manifests itself in two different aspects: the even-to-odd magnitude itself in the first region and its main oscillation frequency in the second region. We note that the predictive power of Eq.~(\ref{eq:e2o}) is optimized [the (i) regime] if the primary laser's parameters vary in an appropriate working range (intensity within $[1.0 -  4.0] \times 10^{14}$~W/cm$^2$ and wavelength longer than 1200~nm (mid-IR laser))~\cite{suppl}. In fact, the primary laser field $E_0$ must be high enough not only to ensure a low controlled ratio $E_T/E_0$ to avoid deforming the HHG plateau structure, but also not exceed the intensity saturation. Also, the wavelength should be long so that the asymmetry parameter $\gamma$ can fall into the stable range.

\textit{Application in THz waveform sampling} --- With laser parameters that optimize the validity of Eq.~(\ref{eq:e2o}), we propose a pump-probe THz waveform sampling scheme, illustrated in Fig.~\ref{fig:extract}(a). Here, the THz pulse becomes the pump pulse, being probed by a delayed mid-IR laser. The two pulses meet inside a gas jet, triggering the HHG process. The odd-even HHG [Fig.~\ref{fig:extract}(ii)] and, consequently, the even-to-odd ratio~[Fig.~\ref{fig:extract}(iii)] are recorded at each time delay. With the analytical formula~(\ref{eq:e2o}), the waveform of the THz pulse can be easily reconstructed [Fig.~\ref{fig:extract}(iii)].

Figure~\ref{fig:extract} shows an example of the waveform measurement for the THz pulse $E_T (t) = E_{T0} \exp\left(-\omega_T^2 t^2/36 \pi^2 \right) \sin \omega_T t$ with $E_T = 257$~kV/cm and frequency $\omega_T = 1.3$~THz, inspired by the pulse recently reported in \cite{Shalaby:ol16}. The five-cycle trapezoidal primary pulses with $2.5 \times 10^{14}$~W/cm$^2$ intensity and 2000~nm wavelength are used as probe lasers. The time resolution of the THz waveform sampling is related to the probe pulse duration, which is about 20~fs for the five-cycle laser pulse with three cycles in the flat part. Figure~\ref{fig:extract}(b) demonstrates the validity of our proposed procedure. It indicates a good consistency between the extracted THz waveform and the input data. We have also examined and affirmed the validity of the proposed method in sampling THz pulses with complicated waveforms implying a broad frequency band (not shown). We note that the carrier-envelope phase of the detected THz pulse might be flipped by $\pi$ since the proposed method can extract the magnitude only but not its sign.

Using the optimized universal rule can detect the THz electric field within a wide THz field range of about $[20,2000]$~kV/cm imposed by the working range of the probing laser pulses~\cite{suppl}. We emphasize that the detectable range can be expanded if we use the suboptimal regime [regime (ii) of Fig.~\ref{fig:class}] of $\gamma$, looking only at its stable spacing between reversal points. We leave the details in future work.

In \textit{conclusion}, we have both numerically and analytically demonstrated the universal dependence of the even-to-odd ratio on the scaled THz electric field (asymmetry parameter of the system). The approach to derive the universal rule in this Letter can be generalized to other asymmetric laser-target systems that may be meaningful in probing atomic/molecular quantum dynamics, controlling electron dynamics within attosecond time scale, or extracting asymmetric factors of laser-target systems.

Based on this universal rule, we have proposed a pump-probe method for THz waveform sampling using the even-to-odd ratio, which is measurable within current compact laser setups. Unlike the previous methods for THz detection involving electronic or optical excitation of targets under THz pulse, our proposed method works on the perturbative regime in which the THz field only affects the dynamics of quasi-free electrons in the continuum energy region leading to modulation of the even-to-odd ratio but does not directly interact with the targets. This independence of the targets and probe laser parameters makes the method feasible for detecting a wide range of THz electric fields.

\begin{acknowledgments}
\textit{Acknowledgments - }
This work was funded by Vingroup and supported by Vingroup Innovation Foundation (VINIF) under project code VINIF.2021.DA00031. The calculations were executed by the high-performance computing cluster at Ho Chi Minh City University of Education, Vietnam.

\end{acknowledgments}


\bibliographystyle{apsrev4-1}
\bibliography{MyBib}
\end{document}